\documentclass[letterpaper,11pt]{article}

\usepackage{jheppub}
\usepackage{natbib}
\usepackage{latexsym}
\usepackage{psfrag}
\usepackage{graphicx}
\usepackage{amsmath}
\usepackage{amssymb}
\usepackage{xspace}
\usepackage{color}
\usepackage{multirow}
\usepackage{colortbl}
\usepackage[]{xcolor}
\usepackage{bbm}
\usepackage{mathtools}

\newcommand{\cO}{\mathcal{O}}
\newcommand{\be}{\begin{equation}}
\newcommand{\ee}{\end{equation}}
\newcommand{\bea}{\begin{eqnarray}}
\newcommand{\eea}{\end{eqnarray}}
\newcommand{\hpc}{H$\pi \chi$PT}
\newcommand{\chpt}{$\chi$PT~}

\newcommand{\mpi}{M_\pi}
\newcommand{\im}{\mathrm{Im\,}}

\newcommand{\bbar}{\overline}

\allowdisplaybreaks[2]

\title{On the factorization of chiral logarithms \\ in the pion form factors}

\author[a]{G.~Colangelo,}
\author[a]{M.~Procura,}
\author[a]{L.~Rothen,}
\author[a]{R.~Stucki,}
\author[b]{J.~Tarr\'us Castell\`a}

\affiliation[a]{Albert Einstein Center for Fundamental Physics, Institute
  for Theoretical Physics, \\ University of Bern, Sidlerstr. 5, CH-3012
  Bern, Switzerland} 
\affiliation[b]{Departament d'Estructura i Constituents de la Mat\`eria and
  Institut de Ci\`encies del Cosmos
Universitat de Barcelona, Diagonal, 647, E-08028 Barcelona,
  Catalonia, Spain \vspace{2ex}}

\emailAdd{gilberto@itp.unibe.ch}
\emailAdd{mprocura@itp.unibe.ch}
\emailAdd{stucki@itp.unibe.ch}
\emailAdd{rothen@itp.unibe.ch}
\emailAdd{tarrus@ecm.ub.es}

\abstract{
The recently proposed hard-pion chiral perturbation theory predicts that the leading
chiral logarithms factorize with respect to the energy dependence in the chiral
limit. This claim has been successfully tested in the pion form factors
up to two loops in chiral perturbation theory. In the present
paper we explain this factorization property at two loops and
even show that it is valid to all orders for a subclass of diagrams. We
also demonstrate that factorization is violated starting at three loops.
}

\begin{document}

\maketitle

\section{Introduction}
\label{sec:intro}
Chiral perturbation theory (\chpt \!)~\cite{Weinberg:1978kz,Gasser:1983yg}
allows one to calculate the momentum and light quark mass dependence of
hadronic processes at low energy. It can also be applied to processes in
which hard scales are involved, {\em e.g.} when baryon or heavy quarks
interact with pions. In these cases, however, the range of applicability of
the theory is restricted to the kinematical regions in which the momenta of
the pions are soft. In ref.~\cite{Flynn:2008tg} Flynn and Sachrajda
proposed an interesting argument which, 
after reinterpreting the effective Lagrangian describing the decay of a
heavy meson (the $K$ in that case) into a pion and a lepton-neutrino pair,
leads to a prediction of the coefficient of the leading chiral logarithm
in $K_{\ell 3}$ decays even for kinematical configurations where the pion
is not soft.
Subsequently, in a series of papers by Bijnens and
collaborators~\cite{Bijnens:2009yr,Bijnens:2010ws,Bijnens:2010jg,Bijnens:2011bp}
it has been claimed that the calculation of such a chiral logarithm is
possible in more general processes in which the pion is hard. This approach
has been referred to as hard pion chiral perturbation theory (\hpc).

A particularly clear example of a prediction in \hpc~has
been provided in ref.~\cite{Bijnens:2010jg} where it has been shown that the
scalar and vector form factors of the pion, which have been calculated to
two loops in $SU(2)$ \chpt \cite{Bijnens:1998fm}, in the limit $M_\pi^2/s \ll 1$
factorize:
\be \label{eq:fact}
F_{V,S}(s,\mpi^2) = \bbar{F}_{V,S}(s)\, \big(1+ \alpha_{V,S}(s) \,L \big) +
{\cal O}(M^2)~. 
\ee 
Here $s$ is the momentum transfer squared and $L$ stands for the leading
chiral logarithm, defined as\,\footnote{Writing $\ln M^2/ s$ as $\ln M^2/
  \mu^2 + \ln \mu^2/ s$, one can then equivalently define $L$ as
  $\mu^2$-dependent as done in refs.~\cite{Flynn:2008tg,Bijnens:2010jg} and
  the second term goes into the $\cO(M^2)$ contribution in
  eq.~(\ref{eq:fact}).}  
\be
L \equiv \frac{M^2}{(4 \pi F)^2}\, \ln \frac{M^2}{s} ~. 
\ee
$M^2$ is proportional to the average up and down quark masses $\hat{m}$
($M^2=2B \hat{m}$) and $F$ is the pion decay constant in the chiral limit: 
\be
\mpi^2= M^2 + \cO(M^4)~, \qquad F_\pi = F + \cO(M^2)~.
\ee
$\bbar{F}_{V,S}(s)$ denote the form factors for vanishing $u$ and $d$ quark
masses\,\footnote{In this paper quantities in the $SU(2)$ chiral limit are 
  denoted by a bar,  $\bbar{X} = \lim_{\mpi \to 0} X$.}. Bijnens and Jemos
provide arguments in support of the validity of this factorization 
property to all orders \cite{Bijnens:2010jg}. On the other hand a precise
formulation of this factorization property is still lacking and it is 
not clear under what circumstances and for what quantities it holds.

The aim of this paper is to start providing an answer to these questions.
In particular we asked ourselves whether factorization of chiral logs for
asymptotically small values of the quark masses could be exact, or in other
words, a property of QCD. If the latter is true, then this property must
emerge also in any effective theory of QCD provided this can deal with the
chiral limit. This is the case of \chpt in the limit $M^2 \ll s \ll
\Lambda_\chi^2$, and indeed the observed factorization in the form factors
at two-loop level points in this direction. In the present paper we will
analyze the pion form factors in \chpt beyond two loops. We find that the
most effective way to attack this problem is to study the dispersion
relation for the form factors and apply the chiral counting to this. As we
will show, this allows us to formulate a recursive argument and to
establish factorization for a subclass of diagrams to all orders in the
chiral counting.

We also find, however, that inelastic contributions (three-loop diagrams
with a four-pion intermediate state) also generate chiral logs of the order
of interest to us and that these violate the factorization property. These
diagrams do not fall in the subclass of diagrams which are treated in our
recursive argument mentioned above. We conclude that factorization of
chiral logs cannot be an exact property of QCD for the pion form factors.
If we believe that \hpc~is valid even for (asymptotically) large values of
$s$, then this conclusion would be implied by the available results on the
meson form factors in the limit of asymptotically large energy. In this
limit the pion form factors can be
analyzed in the context of perturbative QCD factorization, as shown by
Brodsky-Lepage \cite{Lepage:1980fj}. In their formula the quark masses are
neglected, but as discussed by Chen and Stewart \cite{Chen:2003fp}, it is
possible to study the deviation from the chiral limit and in particular to
determine the coefficient of the leading chiral log. This chiral log
factorizes with respect to the leading Brodsky-Lepage term (but there is no
reason to assume that it could factorize for the subleading terms).
However, the coefficient of the leading chiral log is different from what
is obtained in the low-energy limit.

The paper is organized as follows: in sec.~\ref{sec:disp} we introduce our
notation and briefly discuss the dispersive representation of the pion form
factors. In sec.~\ref{sec:elastic} we show how, after applying the chiral
counting, one can determine the leading chiral log on the basis of the
dispersive representation. Considering only elastic contributions (two-pion
intermediate states) we are able to provide a recursive argument and prove
factorization of the leading chiral log to all orders. In
sec.~\ref{sec:inelastic} we discuss the leading inelastic contributions,
calculate the chiral log coming from these and show that these violate
factorization. In sec.~\ref{sec:BLCS} we briefly discuss the papers of
Brodsky-Lepage \cite{Lepage:1980fj} and Chen-Stewart \cite{Chen:2003fp}
and show that these also lead to the same conclusion, namely that
factorization of the leading chiral log cannot be exact.

\section{Dispersive representation of the pion form factors}
\label{sec:disp}
We consider the vector and scalar pion form factors, respectively $F_V(s)$
and $F_S(s)$ and denote them by the common symbol $F(s)$ (unless it is
necessary to distinguish them). We normalize the scalar form factor 
such that\,\footnote{Notice that, consistent with this normalization, we define
leading order as $\cO(p^0)$, next-to-leading order as $\cO(p^2)$ and so
on. The reader should be aware of the mismatch with the usual counting for
the Lagrangian and the $\pi \pi$ scattering amplitude discussed below,
where leading order is $\cO(p^2)$.}  $F_S(0)=1$ --- for the vector form factor this condition follows
from current conservation. Note that this normalization condition (whether
automatic or imposed) is inessential to our argument. Both these form
factors are analytic functions in the cut plane $[4 \mpi^2, \infty)$ and
satisfy a once subtracted dispersion relation of the form
\be
\label{eq:DR}
F(s) = 1 + \frac{s}{\pi} \int_{4 \mpi^2}^\infty ds'\, \frac{\im
  F(s')}{s'(s'-s)}~.
\ee
The subtraction constant is the value of the form factor at zero momentum
transfer which is equal to one as mentioned above. In the elastic region
unitarity relates the imaginary part of the form factor to the form factor
itself and the $\pi \pi$ partial wave with the same quantum numbers: 
\be
\label{eq:imF}
\im F(s) = \sigma(s)\, F(s)\, t^*(s)~, \qquad \sigma(s) = \sqrt{1-\frac{4
    \mpi^2 }{s}}~. 
\ee
When $s$ gets larger than the inelastic threshold, additional cuts
involving several intermediate pions contribute to the discontinuity. In
the subsequent discussion we will consider the form factor which is
obtained by solving the dispersion relation with imaginary parts arising
only from two-pion intermediate states. We will be able to make a statement
about this contribution, provided we consider only two-pion intermediate
states also for the $\pi \pi$ scattering amplitude entering the unitarity
relation (\ref{eq:imF}). We call this part the elastic contribution, and
the rest the ``inelastic'' part. 
\be
F(s)=F_{\mathrm{el}}(s)+F_{\mathrm{inel}}(s)~.  
\ee 
We stress that this definition is a diagrammatic one. Its precise
meaning should become more clear in the following section, in which we will
consider only the elastic contribution. The first diagrams contributing to
$F_{\mathrm{inel}}(s)$, which start at order $p^6$, are discussed in
sec.~\ref{sec:inelastic}. 

In the following we will apply the chiral counting to the dispersion
relation \cite{Gasser:1990bv,Colangelo:1996hs}.  It is
well-known\footnote{For a pedagogical discussion see
  \cite{Donoghue:1995pd}.} that if one does so one gets a one-to-one 
correspondence between direct \chpt calculations and the various chiral
orders in the dispersion relation: in other words \chpt provides a
perturbative solution to the dispersion relation. This will be the basis of
our analysis\,\footnote{Early calculations of the chiral logs in chiral perturbation theory have been made
following a similar approach \cite{Gasser:1979hf,Gasser:1980sb}.}. Indeed, since $t(s)$ starts at $\cO(p^2)$, knowing the form
factor and the $\pi \pi$ partial wave up to a given chiral order means to
know $\im F(s)$ at one order higher. If one is then able to perform the
dispersive integral, the corresponding real part of the form factor is also
obtained.  This is a powerful tool which allows us to argue recursively.

The chiral representation contains of course more information than what one
gets from the dispersion relation as it also gives the quark mass
dependence of subtraction constants. Indeed whenever necessary we will use
\chpt to determine chiral logs in subtraction constants. This will be
needed in the $\pi \pi$ scattering amplitude, but not for the form factors
since with the normalization condition we chose for the form factors
(\ref{eq:DR}), the subtraction constant is equal to one and cannot contain
any logs.

\section{The elastic contribution to $F(s)$}
\label{sec:elastic}
We will now address the question, how leading chiral logarithms can arise
from the elastic contribution to the dispersive integral. There are two
possible mechanisms to generate terms containing $L$: either one starts
from an integrand which does not contain a log of the pion mass and this is
produced by the integration over $s'$, or the integrand itself contains a
chiral log and the dispersive integral determines what function of $s$
multiplies it. We will now discuss these two mechanisms in turn.  We stress
that, as mentioned in the introduction, we are considering the form factors
in the limit $M^2 \ll s$ and are only interested in terms proportional to
$L$: terms of $\cO(M^2)$ without logarithmic enhancement or terms of
$\cO(M^2 L)$ are all beyond the accuracy we aim at.

To avoid clutter in the notation in this section we will drop the
subscript ``el'', $F(s)\mathrel{\widehat{=}} F_{\mathrm{el}}(s)$, as this
is the only contribution discussed here.

\subsection{Leading chiral logarithms from the dispersive integration} 
\label{sec:integration} 

The first possible mechanism is that the leading chiral logarithms are
generated by the dispersive integral, i.e. they are produced at the lower
integration boundary $s'\sim 4 \mpi^2$, which goes to zero in the chiral
limit. In order to investigate this, we must analyze the behavior of the
integrand in the regime $s' \sim M^2 \ll s$: we can therefore make use of
the standard chiral expansion for the discontinuity of the form factor
(\ref{eq:imF}).  Expanding both the form factor and the $\pi \pi$
amplitude we can write
\be
\label{eq:t_new}
F(s, \mpi^2) = 1 + \cO(s)+ \cO(M^2) \qquad \mbox{and} \qquad
t(s, \mpi^2) = c_1\, M^2 + c_2\, s + \cO(s^2)+ \cO(M^4)~. 
\ee 
Plugging eq.~(\ref{eq:t_new}) into eq.~(\ref{eq:DR})
we obtain 
\begin{equation} 
\label{eq:disp} 
  F(s) = 1 + \frac{s}{\pi} \int_{4 M_{\pi}^2}^{\infty} ds' \frac{\sigma 
    (s')}{s'(s'-s)} \, \Big(c_1 \,M^2+c_2\,s' +\mathcal{O}(p^4)\Big)~.  
\end{equation}

The three terms in brackets in the integrand in eq.~(\ref{eq:disp})
generate three types of integrals: the first (the one proportional to 
$c_1$) is the well-known loop function $\bar{J}(s)$ which has the following
expansion in $M^2/s \ll 1$,   
\begin{equation} 
\label{eq:jbar} 
\frac{s}{\pi} \int_{4 M_{\pi}^2}^\infty ds' \frac{\sigma (s')}{s'(s'-s)} = 
16 \pi \,\bar{J}(s) =  \frac{1}{ \pi} \left[1+ \ln  
  \frac{M^2}{-s} + \frac{2M^2}{s} \left(1 - \ln 
 \frac{M^2}{-s}\right)  +  
  \cO\left(\frac{M^4}{s^2}\right) \right]~.  
\end{equation} 

The remaining two integrals are UV divergent. However, since we are 
interested in their behaviour close to the lower integration boundary, we 
can introduce an $\mpi$-independent cut-off $\Lambda^2$, which allows 
us to interchange integration and expansion for small $M$. The second type 
of integral (the one proportional to $c_2$) is then given by 
\be 
\frac{s}{\pi} \int_{4 \mpi^2}^{\Lambda^2} ds' \frac{\sigma (s')}{(s'-s)} = 
\bbar{d}_1(s, \Lambda^2)-\frac{2}{\pi}\,M^2 \ln \frac{M^2}{s}+\cO(M^2)  
\ee 
and the third by 
\be 
\frac{s}{\pi} \int_{4 \mpi^2}^{\Lambda^2} ds' \frac{\sigma (s')}{s'(s'-s)} 
\times \cO(p^4)= \bbar{d}_2(s, \Lambda^2)+\cO(M^2)~.  
\ee
With $\cO(p^4)$ we indicate terms proportional to $s'^2$, $s' M^2$ and 
$M^4$, which are all small in the region close to the lower integration 
boundary. It is easy to see that this further suppression does not allow 
terms proportional to $L$ (without further powers of $M^2$) to be 
generated. We conclude that the dispersive integral can generate leading 
chiral logs only from the leading chiral contribution to the integrand.
The functions $\bbar{d}_1(s, \Lambda^2)$ and $\bbar{d}_2(s, \Lambda^2)$ are
quark mass independent and contribute to $\bbar{F}(s)$ together with the 
low-energy constants (LECs) which cancel the $\Lambda$-dependence. 

Putting all pieces together we conclude that the chiral log generated by
the dispersive integration is given by
\be
16 \pi F^2 \left(c_1- 2\, c_2 \right) L~. 
\ee
The constants $c_1$ and $c_2$ are related to the leading chiral 
contributions to the $\pi \pi$ scattering lengths and effective ranges 
characterizing the threshold expansion  
\be 
\label{eq:t_threshold} 
{\rm Re}\,t_\ell^I(s) = q^{2 \ell}\left(a^I_\ell + b^I_\ell q^2 + \cO(q^4) 
\right)~,
\ee
where $q^2= s/4 - \mpi^2$ and $I$ and $\ell$ denote isospin and angular 
momentum, respectively. For the scalar form factor the relevant parameters 
are $c_1=a^0_0/M^2 - b_0^0 + \cO(M^2)$ and $c_2=b^0_0/4+ \cO(M^2)$ 
where~\cite{Gasser:1983yg}  
\be 
a^0_0= \frac{7 M^2}{32 \pi F^2} +\cO(M^4) \; , \qquad b_0^0= \frac{1}{4 \pi
  F^2} +\cO(M^2)~, 
\ee 
leading to
\be 
 \alpha_S=16 \pi F^2 \left(c_1- 2\, c_2 \right)=16 \pi F^2 \lim_{M^2 \to 0}
 \left(\frac{a^0_0}{M^2}-\frac{3b^0_0}{2} \right)=-\frac{5}{2}~,
\label{eq:alphaS}
\ee 
which reproduces the known result
\cite{Bijnens:1998fm,Bijnens:2010jg}. For the vector form factor we 
must instead use the parameters: $c_1=- (a^1_1)_{|_{M^2 \to 0}}=-1/(24 \pi 
F^2)$, and $c_2=(a^1_1/4)_{|_{M^2 \to 0}}=1/(96 \pi F^2)$, which leads to 
\be 
 \alpha_V= 16 \pi F^2 \left(c_1- 2\, c_2 \right)=16 \pi F^2 \lim_{M^2 \to
   0} \left(-\frac{3 a^1_1}{2} \right) =-1~, 
\label{eq:alphaV} 
\ee 
also in agreement with the explicit calculation in 
refs.~\cite{Bijnens:1998fm,Bijnens:2010jg}.   

For the subsequent discussion it is useful to determine the leading order 
of $\bbar{F}(s)$ in the chiral expansion in powers of $s$ (see also 
\cite{Bissegger:2006ix}). In order to do this, we have to find a function
which is analytic in the cut plane $[0,\infty)$ and with the following
imaginary part along the cut  
\be 
\lim_{M\to 0}  \left[\sigma(s) F^{(0)}(s)\, t^{(2)}(s) \right]=c_2\, s~. 
\ee 
Such a function is easily found: 
\be 
\bbar{F}^{(2)}(s) = s\, \frac{c_2}{\pi} \ln \frac{\Lambda_2^2}{-s}~, 
\label{eq:F2chi} 
\ee 
with $\Lambda_2$ an unknown scale. The explicit expressions in the case of 
the scalar and vector form factors read 
\bea 
\bbar{F}_S^{(2)}(s) &=& \frac{s}{16 \pi^2 F^2} \left[1+\ln \frac{\mu^2}{s}
  + i \pi + 16 \pi^2 \ell_4^r(\mu) \right] \nonumber \\
\bbar{F}_V^{(2)}(s) &=& \frac{s}{16 \pi^2 F^2} \left[\frac{5}{18}+ 
  \frac{1}{6}\ln \frac{\mu^2}{s} + \frac{i \pi}{6} - 16 \pi^2 \ell_6^r(\mu)
\right]~, 
\label{eq:d2bar} 
\eea 
in agreement with ref.~\cite{Bijnens:2010jg}. 
The coefficients of the logarithms are indeed correctly reproduced by 
substituting $c_2=(b_0^0/4)_{|_{M^2 \to 0}}=1/(16 \pi F^2)$ for the scalar
form factor and $c_2=(a_1^1/4)_{|_{M^2 \to 0}}=1/(96 \pi F^2)$ for the
vector one.

\subsection{Leading chiral logarithms from the integrand}
\label{sec:integrand} 

We shall now discuss whether and how leading chiral logarithms can be
generated at higher chiral orders --- more specifically, we are interested
in terms proportional to $s^{n-1} L$ at order $p^{2n}$.  The discussion
above made it clear that the behavior of the integrand around the lower
limit of integration may only generate a chiral logarithm at $\cO(p^2)$.
There is a second mechanism, however, by which chiral logarithms may arise
from the dispersive integrals at higher orders, namely if the integrand
itself contains a chiral logarithm.

Let us consider the dispersion relation at $\cO(p^4)$ (i.e. the
contributions to the form factor at the two-loop level). At this order the
integrand has this form
\be \label{eq:ImF4}
\im F^{(4)}(s) = \sigma(s) \left[ t^{(4)*}(s)+F^{(2)}(s)\, t^{(2)}(s) \right]
\ee
and each of the terms may contain chiral logarithms. We consider first the
latter one: 
$t^{(2)}(s)$ is the tree-level contribution to the $\pi \pi$ scattering
amplitude and contains no chiral logarithms, whereas $F^{(2)}(s)$ does, as
we have seen above. Expanding in $M^2/s$, we can write it as
\be
F^{(2)}(s)\, t^{(2)}(s) = \left(\bbar{F}^{(2)}(s) + \alpha L
\right) \bbar{t}^{(2)}(s) + \cO(M^2)~.
\ee
We can similarly expand $t^{(4)}(s)$,
\be \label{eq:beta}
t^{(4)}(s) = \bbar{t}^{(4)}(s)+ \beta\, s L + \cO(M^2)~. 
\ee 
It is easy to realize that the term proportional to $\beta$ would destroy
factorization: this vanishes, however, as shown in
app.~\ref{app:partwav}. We therefore conclude that the only term containing
$L$ in eq.~(\ref{eq:ImF4}) has as coefficient exactly the absorptive part of
the form factor at one chiral order lower times $\alpha$. As we have
discussed above, the solution of the corresponding dispersion relation
reads 
\be
\alpha\,s\, \frac{c_2}{\pi} \,\ln \frac{\Lambda_X^2}{-s}
\ee
with $\Lambda_X$ an unknown energy scale. We argue, however, that this has
to coincide with $\Lambda_2$ introduced in eq.~(\ref{eq:F2chi}). From the
point of view of \chpt both of these scales are related to LECs in the $\cO(p^4)$
Lagrangian: $\Lambda_2$ in particular is the one which appears in the form
factor at this order. Since we have just showed that one cannot generate a
chiral logarithm by integrating over a local contribution of $\cO(p^4)$ to the
$\pi \pi$ scattering amplitude (which is the only other vertex in the
dispersion relation), we have to conclude that $\Lambda_X=\Lambda_2$.
We conclude that at two loops we can write the contribution to the form
factor as  
\be \label{eq:factel}
F(s)=\left(1+\bbar{F}^{(2)}(s)\right) (1+ \alpha L) + \bbar{F}^{(4)}(s) +
\cO(M^2) + \cO(p^6)~.
\ee
i.e. in a factorized form, as predicted by \hpc.

For the elastic contribution to $F(s)$, the same reasoning can be repeated
in exactly the same way order by order. At every new step the terms
threatening factorization are the contributions to $\im F^{(2 n)}$ arising
from the $\pi \pi$ scattering amplitude at the same order. A chiral
logarithm of the form $s^{n-1} L$ in $t^{(2 n)}$ would destroy factorization.
In app.~\ref{app:partwav} we show that these terms are absent.  However, at
order $p^6$, the four-pion intermediate states contribute to $\im F(s)$ and
we will now show that these yield leading chiral logarithms and are
responsible for the breakdown of factorization, leading to the three-loop
result
\be
\label{eq:inel_fact} F(s)=\left(1+\bbar{F}^{(2)}(s)+\bbar{F}^{(4)}(s)
\right) (1+ \alpha L) + \alpha_{\rm inel}(s) L+ \bbar{F}^{(6)}(s) +
\cO(M^2) + \cO(p^8)~.  \ee

Before closing the section we comment on the form factor in the chiral 
limit: at $\cO(p^4)$ this is given by the solution of the
dispersion relation with discontinuity  
\be
\im \bbar{F}^{(4)}(s) = \left[ \bbar{t}^{(4)*}(s)+\bbar{F}^{(2)}(s)
  \, \bbar{t}^{(2)}(s) \right]~. 
\ee
The form factor $\bbar{F}^{(4)}(s)$ can be derived from
ref.~\cite{Bijnens:1998fm} and given in explicit analytic form. The
expression for the scalar form factor, which will be used for the numerical
analysis in sec. 4, {\em e.g.} reads
\be
\bbar{F}_S^{(4)}(s)= \frac{s^2}{F^4}
\left[\frac{43}{36} \tilde{L}^2-\tilde{L}\left(\frac{1}{36N}+ \ell_a^r
\right)
+\frac{1}{9N}\left(\frac{7}{192}+(\ell_1^r+2\ell_2^r)+\frac{305}{144 N}
\right)  + r_{S 3}^r \right]
\ee
where $N:=16 \pi^2$, $\ell_a^r:=1/3(11\ell_1^r+7\ell_2^r)+\ell_4^r$ and 
$\tilde{L}:= 1/N \left(\ln(s/\mu^2)- i \pi -1 \right)$. The $r_{S 3}^r$ LEC
stems from the $\cO(p^6)$ chiral Lagrangian.

\section{The contribution from inelastic channels}
\label{sec:inelastic}
In the previous section we have shown that the leading chiral logarithms
originating from the elastic part factorize. We are now going to show that
the $\cO(p^6)$ inelastic contribution leads also to terms proportional to
$s^2\,L$, i.e. $\alpha_{\rm inel}(s) \neq 0$ in eq.~(\ref{eq:inel_fact}),
which implies that the factorization hypothesis at the basis of \hpc~is not
valid to all orders. We shall now present the details of our calculation.

The lowest order inelastic contribution to $F(s)$ is given by three-loop
diagrams with four intermediate pions. We evaluated them by means of the
following dispersion relation with the lower integration boundary given by
the four-pion threshold: 
\be
\label{eq:4pion}
F_{\mathrm{inel}}(s)=\frac{s}{\pi}\int_{16 M_\pi^2}^\infty
ds'\,\frac{\mathrm{Im}\,F_{\mathrm{inel}}(s')}{s'(s'-s)}~.   
\ee  
Chiral logarithms are either produced by the dispersive integration or are
contained in the integrand. In the chiral counting $\im F_{\mathrm{inel}}$
is of $\cO(p^6)$ and contains a four-particle phase space factor which
gives a strong suppression near threshold ( $\Phi_4(s) \sim (s-16
M_\pi^2)^{7/2}$). Arguing similarly to how we did in
sec.~\ref{sec:integration} we conclude that leading chiral logs cannot be
generated by the dispersive integration. Therefore we need to concentrate
on the integrand.

\begin{figure}[t] 
 \centering 
 \includegraphics[width=1\textwidth,height=0.2\textheight]{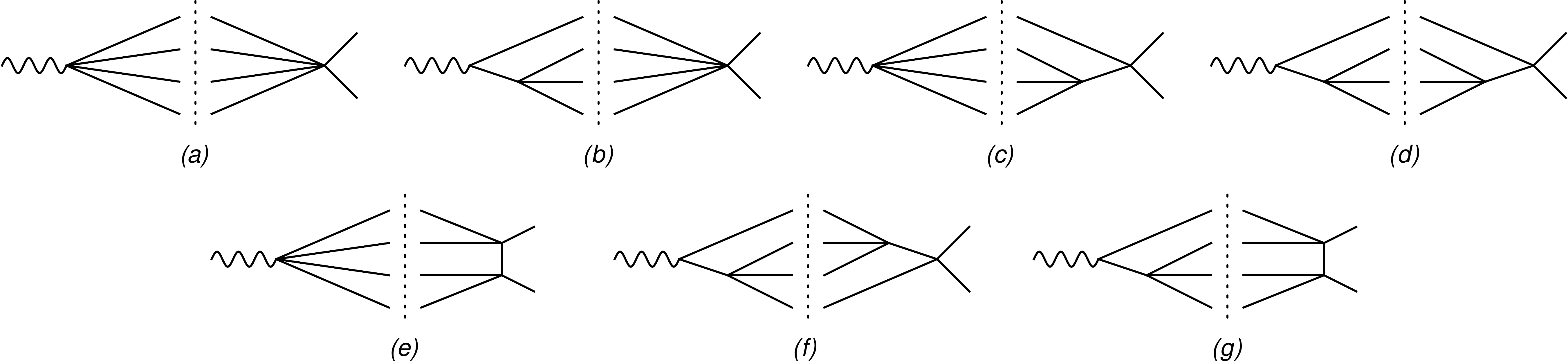}  
 \caption{Three-loop cut diagrams contributing to the scalar
   $F_{\mathrm{inel}}(s)$ at order $p^6$.} 
 \label{img:3loop}
\end{figure}

Due to unitarity, the imaginary part of the form factor has this form  
\be 
\label{eq:ImFin} 
\mathrm{Im}\,F_{\mathrm{inel}}(s)=\frac{1}{2}\int
d\Phi_4(s;p_1,p_2,p_3,p_4)\, F_{4\pi}\cdot T_{6\pi}^*~,  
\ee 
where $d\Phi_4$ is the phase space for four pions of momenta $p_1, \dots, 
p_4$. $F_{4\pi}$ denotes the current-$4\pi$ vertex and $T_{6\pi}$ is the 
six-pion scattering amplitude. The three-loop diagrams
contributing to 
$F_{\mathrm{inel}}^{(6)}(s)$ are shown in fig.~\ref{img:3loop}.   

Chiral logarithms are produced by integrations over intermediate momenta
with mass-dependent boundaries. We find that, in order to calculate the
terms proportional to $L$, we can expand the integrand for small $M$ and
keep only the relevant terms. It is then crucial to find a phase space
parametrization which allows us to perform analytical integrations after
this expansion. We illustrate here a suitable choice for $d\Phi_4$.

We reduce the $4$-body phase space defined as 
\be
\label{eq:4body_form} 
d\Phi_4(s;p_1,...,p_4)=\left(2 \pi\right)^4\delta^4\left(P-\sum_{i=1}^{4} 
  p_i\right)\prod_{i=1}^{4}\frac{d^3 \vec{p}_i}{(2\pi)^3 2 p_i^0}~.  
\ee 
with $s=P^2$ to the following product of two-body phase space factors (see for example sec. 43 of ref.  \cite{Beringer:1900zz})
\be
\label{eq:4body} 
d\Phi_4(s;p_1,p_2,p_3,p_4)=d\Phi_2(s;q,p_4)\times d\Phi_2(q^2;k,p_3)\times 
d\Phi_2(k^2;p_1,p_2)\frac{dk^2}{2\pi}\frac{dq^2}{2\pi}~,  
\ee 
where  
\be 
k=p_1+p_2\,\,,\,\,\,\,\,q=p_1+p_2+p_3\,\,,\,\,\,\,\,
p_1^2=p_2^2=p_3^2=p_4^2=\mpi^2~. 
\ee 
This gives  
\be 
 \label{eq:4pionPS} 
d\Phi_4= 
\,\frac{1}{(4\pi)^6}\frac{\lambda^{1/2}(s,q^2,\mpi^2)}{2s}\, 
\frac{\lambda^{1/2}(q^2,k^2,\mpi^2)}{2q^2}\,\frac{\lambda^{1/2}(k^2,\mpi^2,\mpi^2)}{2k^2}\,d
\Omega_k\,d \Omega_q\,d \Omega_s\,\frac{dq^2}{2\pi}\frac{dk^2}{2\pi} 
\ee
in terms of the well-known K\"all\'en function $\lambda(x,y,z) \equiv x^2 +
y^2+z^2-2(xy+xz+yz)$. $\Omega_s$ is the solid angle spanned by the unit 
vector $\hat{q} \equiv \vec{q}/|\vec{q}|$ in the center-of-mass 
frame of the two final pions. $\Omega_q$ is the solid angle spanned by 
$\hat{k}$ in the frame where $\vec{q}=0$ and $\Omega_k$ is the solid angle 
spanned by $\hat{p}_1$ in the frame where $\vec{k}=0$. More details on this
phase space parametrization are given in app.~\ref{app:Lorentz}.  
The advantage of using this representation is that each $\lambda$ function 
contains $\mpi^2$ as an argument, which enables us to expand all factors 
and perform all the integrations analytically, at least for the diagrams 
$(a)$, $(b)$, $(c)$ and $(d)$.  

The integration range for the kinematic variables $k^2$ and $q^2$ is 
determined by the delta function which ensures momentum conservation:  
\be 
4\mpi^2 \le k^2 \le (\sqrt{q^2}-\mpi)^2\,\,, \,\,\,\,\,9\mpi^2 \le q^2 \le 
(\sqrt{s}-\mpi)^2~.  
\ee 
We stress that chiral logarithms stem from both upper and lower integration
boundaries due to the functional form of the phase space.  

Let us now consider the scalar form factor. For the diagrams $(a)$ to $(d)$
we have been able to determine analytically both the values in the chiral
limit and the coefficients of $L$ for $\im F(s)$. The latter ones are given
by  
\be
\delta_i\, \frac{s^2}{(4 \pi F)^4}
\ee
where the numerical factors are 
\be
\delta_a=-\frac{10}{3} \pi, \quad \delta_b=\frac{55}{12} \pi,  \quad
\delta_c=\frac{25}{6} \pi,  \quad \delta_d=-\frac{9}{4} \pi 
\ee
For the remaining diagrams the structure is too complicated to perform all
integrations analytically. 
In order to extract the coefficient of the leading chiral logarithm from
the remaining diagrams we set up the following procedure. We performed
analytical integrations whenever it was possible and the remaining
integrations were done numerically. For each diagram we then generated a
large number of points within a certain range of values of $M^2/s$ and
fitted the values of the amplitude minus its chiral limit value with a
functional form dictated by \chpt\!. Our optimal choice for the number of
points, the range in $M^2/s$ and the truncation in powers of $M$ and $s$ of
the fit function was determined by our ability to reproduce the
coefficients $\delta_i$ for the diagrams $(a)$ to $(d)$ within one per
cent.  

Summing up the contributions from all seven graphs we obtain the
coefficient of $L$ in $\im F_{\rm inel}(s)$ at three loops: 
\be
\delta\, \frac{s^2}{(4 \pi F)^4} \quad {\rm with}\quad \delta=(-0.53 \pm
0.05) \, \pi 
\label{eq:deltatot}
\ee
The error on $\delta$ takes into account the uncertainties both in the
numerical integration and in the fit. The pion mass dependence resulting
from the sum of the seven diagrams is definitely not compatible with a
vanishing coefficient for the chiral logarithm. As a check, we tried to
fit with a high degree polynomial in $M^2$ but without the leading chiral
log and obtained unacceptably high values of $\chi^2$.

We checked our results also using the four-body phase-space
parametrization in ref.~\cite{Guo:2011ir}. This is not useful to do
analytical integrations but it provides an important check of our numerical
routine since it involves different kinematic variables and angles compared
to our parametrization.

\begin{figure}[t] 
\centering 
\includegraphics[width=0.48\textwidth]{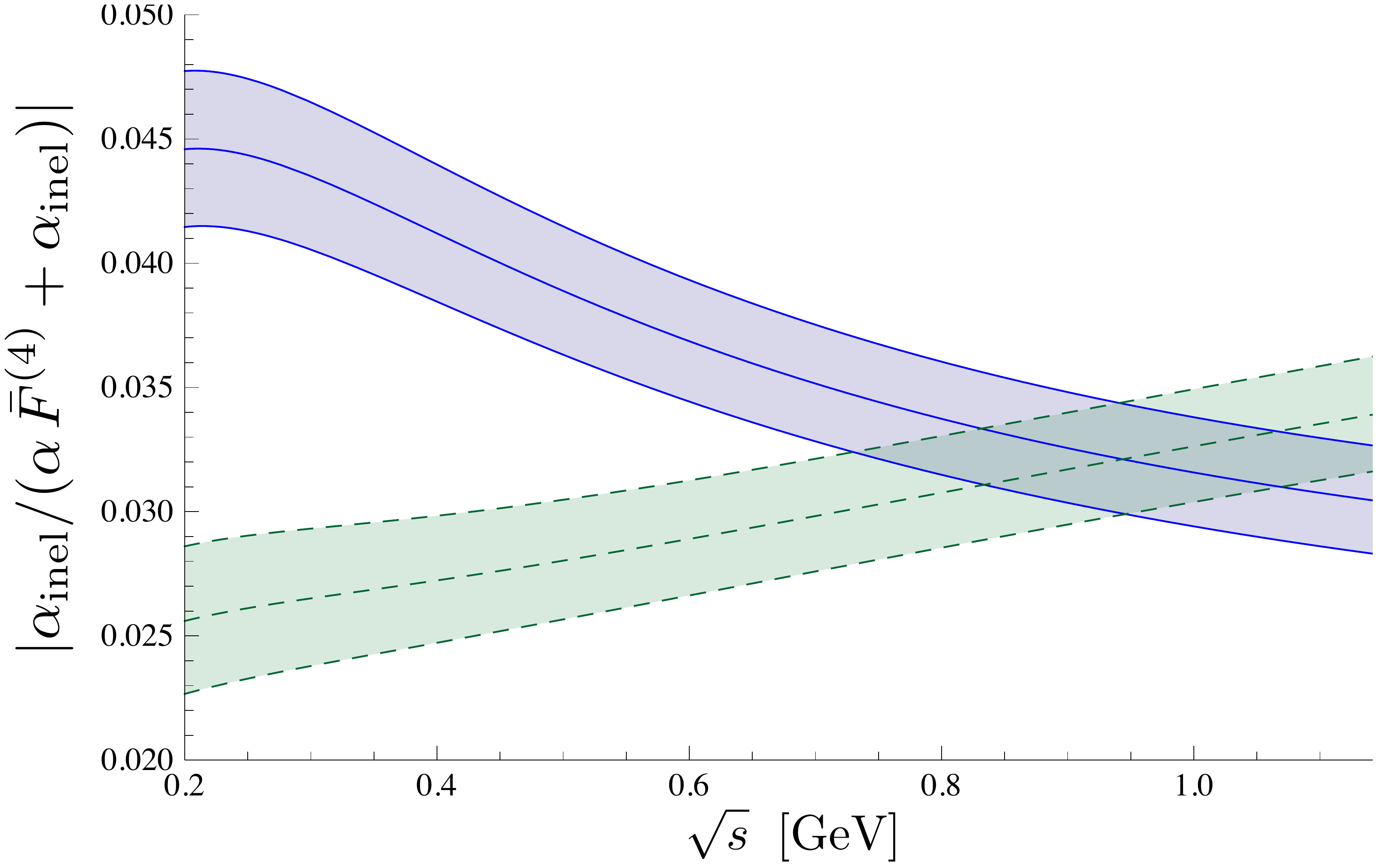} \quad 
\includegraphics[width=0.48\textwidth]{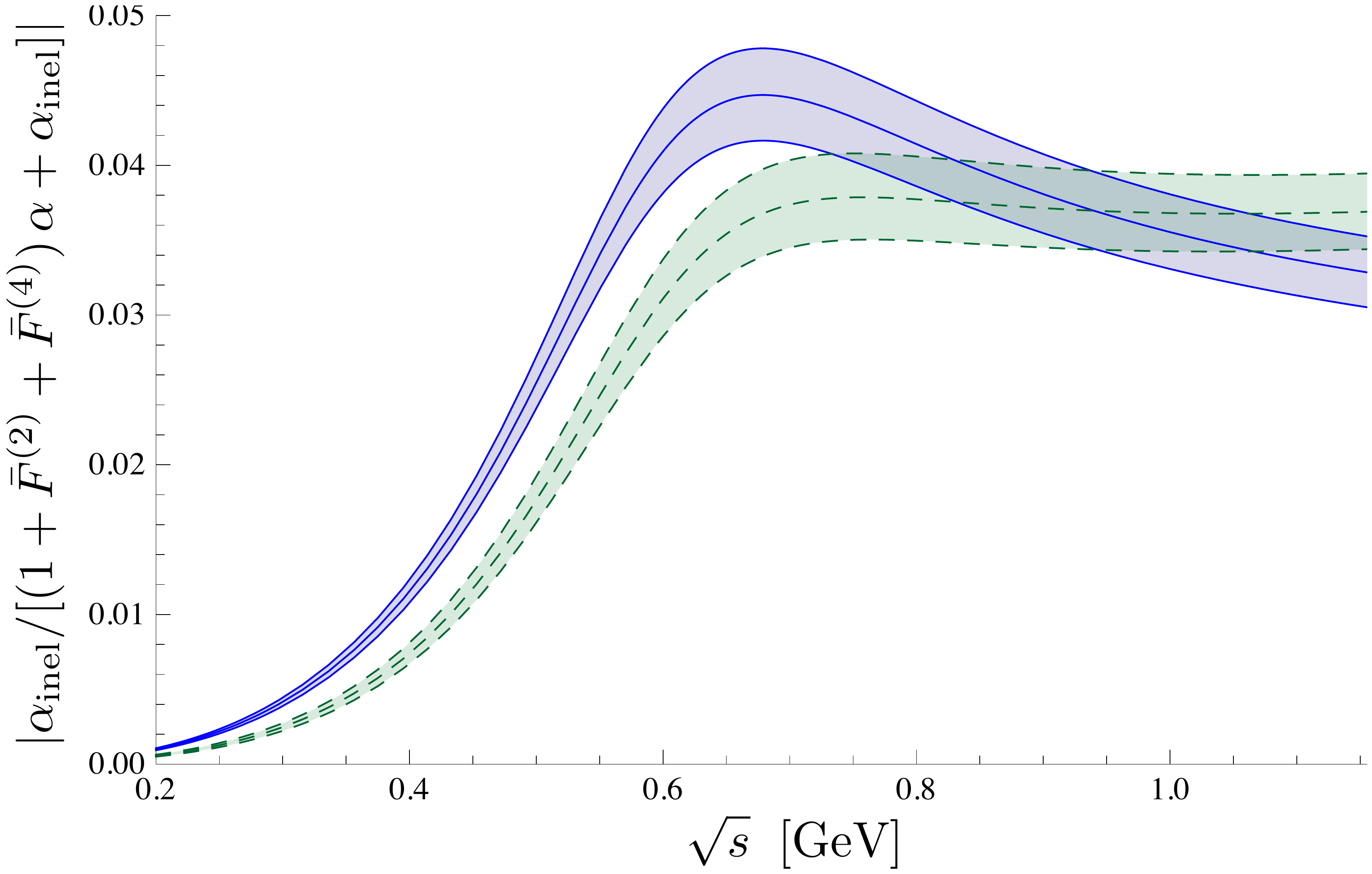} 
\caption{The relative contribution of the inelastic log. The solid line
  corresponds to $C(\mu=1\,{\rm GeV})=1$ in eq.~(\ref{eq:a_inel}) while the
  dashed line to $C(\mu=1\,{\rm GeV})=-1$. The bands corresponds to the
  uncertainty on $\delta$ quoted in eq.~(\ref{eq:deltatot}).} 
\label{fig:corr}
\end{figure}

After performing the dispersive integral in eq.~(\ref{eq:4pion}), our
result for $\alpha_{\rm inel}$ is 
\be \label{eq:a_inel}
\alpha_{\rm inel}(s) = \left[ C(\mu^2) + \frac{\delta}{\pi} \times \left(
    \ln{\frac{\mu^2}{s}} + i \pi \right)  \right] \frac{s^2}{(4 \pi F)^4}
\ee
where $C(\mu^2)$ is a combination of $\cO(p^8)$ LECs which compensates the
$\mu$-dependence. Assuming that $C(\mu = 1\,{\rm GeV})$ is of natural size,
comparing the logarithms from the elastic and inelastic part at three
loops, we find that the factorization breaking effect is about 5\% in the
range of interest for $s$. In fig.~\ref{fig:corr} we plot the relative
contribution of the inelastic log to the total three-loop log (left panel)
and the relative contribution of the inelastic log to the sum of the
elastic and inelastic log up to three loops (right panel). The LECs are set
equal to the values adopted in the analysis of
ref.~\cite{Bijnens:1998fm}. We checked that varying them within their
phenomenological ranges does not change our conclusions. Comparing the two
plots in fig.~\ref{fig:corr}, one can see that for values of $\sqrt{s}$
which are not small compared to $\Lambda_\chi$ the three-loop result is not
suppressed compared to the two-loop and one-loop results, which is a clear
signal of breakdown of chiral power counting. 

If we go beyond the three-loop level, but still use the chiral counting as
a guide, we can write $\alpha_{\rm inel}(s)$ as the following series  
\be
\alpha_{\rm inel}(s) = \sum_{n=2}^{\infty} \left(\frac{\sqrt{s}}{4 \pi F}
\right)^{2 n} c^{(2 n)}(s) 
\label{eq:inelseries} 
\ee
where $c^{(2 n)}(s)$ contains $\ln s/\mu^2$. Our calculation shows that 
$c^{(4)} \neq 0$, which indicates that
in general $c^{(2n)} \neq 0$ for any $n$. At order $p^6$ the inelastic
contribution to the chiral log is a small, but non-negligible correction to
the factorized one. As long as one remains in the chiral regime $M^2 \ll s
\ll \Lambda_\chi^2$, factorization of the chiral log does emerge as a
property which holds to a good approximation. If, on the other hand we
abandon the low-energy region, the chiral counting is not valid anymore
and it becomes impossible to even estimate the relative size of the
inelastic to the elastic contribution since one should sum the whole series
(\ref{eq:inelseries}) and the analogous one for the elastic part, before
being able to draw any conclusion. In fact, if we go to asymptotically
large energies, factorization of the chiral log does emerge again as an
approximate property, but the origin of it changes completely and the
coefficient of the chiral log gets modified, as we are going to discuss in
the next section. Evidently, as the energy increases, the inelastic
contributions become more and more important and factorization does not
hold anymore, until one reaches very high energies where it shows up again
in a very different form.

We also stress that at $\cO(p^8)$ in the form factors there will be
contributions of four-pion intermediate cuts in the $\pi \pi$ scattering
amplitude $t^{(8)}$. These can be additional sources of logarithms.

\section{Chiral logs for asymptotic energies} 
\label{sec:BLCS}
We point out that the factorized form of $F(s)$ conjectured in \hpc,
eq.~(\ref{eq:fact}), is also not consistent with QCD factorization for
asymptotically large values of $s$. In this regime it is
well-known~\cite{Lepage:1980fj, Efremov:1979qk, Chernyak:1983ej} that the
pion electromagnetic form factor can be written as the following
convolution,
\begin{align} \label{bl} 
F_V(s) = \frac{F_\pi^2}{s} \int_0^1 \!dx\, dy\, T(x,y,s)\, \phi_\pi(x) 
\phi_\pi(y) \times [1+ {\cal O}(\Lambda_{\rm QCD}^2/s, M_\pi^2/s)]~.  
\end{align} 
Here $T$ is a hard scattering kernel which can be calculated in
perturbation theory and $\phi_\pi$ is the non-perturbative light-cone
distribution amplitude (LCDA) for the pion. In the language of
Soft-Collinear Effective Theory, this corresponds to the matching of the
electromagnetic current onto the leading effective operator built with four
collinear quark fields, which has a non-vanishing overlap with symmetric
pion states (i.e. containing only energetic modes)~\cite{Bauer:2002nz}.

In ref.~\cite{Chen:2003fp} Chen and Stewart studied the chiral expansion of
the vacuum-to-pion matrix element of the bilocal quark field operator
defining the $\phi_\pi(x)$ in eq.~(\ref{bl}).  They considered the tower of
local axial twist-2 operators $O_k^{A,a}$ related to moments of the pion
LCDA:
\begin{eqnarray}  \label{me}  
&& \langle \pi^{b}|O_{k}^{A,a}|0\rangle  
  = -i F_{\pi}\delta ^{ab}(n\!\cdot\!p_{\pi})^{k+1}
  \int_{0}^{1}\!dx\,\big(1-2x\big)^{k}\,\phi _{\pi}(x)~, 
\end{eqnarray}  
where $n$ is a light-like vector, and proved that for each of these matrix 
elements the leading chiral log is given just by the one in $F_\pi$. 
Therefore, according to eq.~(\ref{bl}) and the result by Chen and Stewart, 
the leading chiral log in $F_V(s)$ does factorize for $s \gg \Lambda_{\rm 
  QCD}^2$ but the coefficient $\alpha_V$ in eq.~(\ref{eq:fact}) is $-2$, 
not $-1$ as predicted by \hpc.

\section{Conclusions} 
\label{sec:concl} 

In this paper we have scrutinized the foundations of hard pion chiral
perturbation theory. We did so on the basis of one example, the pion form
factors: in this, as well as in other quantities, the leading chiral logs
are predicted to factorize in the limit $M^2 \ll s$. For the form factors,
however, this property has been successfully tested by an explicit two-loop
calculation in \chpt \!. Our aim was to investigate whether this
factorization holds even at higher orders in \chpt \!. Since going beyond
two loops within a diagrammatic approach in \chpt is prohibitive, we have
based our analysis on dispersion relations and unitarity, and have
systematically applied the chiral counting to the dispersive
representation. This approach lends itself to a recursive kind of analysis
and allowed us to explain why factorization emerges at the two-loop level.
Moreover, we have been able to show that for a whole subclass of diagrams
this property holds to all orders.

On the other hand, if one considers multipion contributions to the
discontinuity of the pion form factor (in short: inelastic contributions),
one can see that these also generate chiral logs. We have calculated the
coefficient of the leading chiral log in these diagrams at three loops and
shown explicitly that these do not factorize. Factorization of the leading
chiral logs in the hard pion regime is therefore not an exact property. As
long as one remains in the low-energy regime, but takes the quark masses
very small ($M^2 \ll s$) factorization of the chiral logs is valid to a
good approximation, as our numerical analysis has shown. As the energy
increases, however, the inelastic contributions appear to gradually become
more important, until factorization effectively disappears. We conjecture
this also because the behaviour of the form factors for asymptotically
large energies is known. As Chen and Stewart \cite{Chen:2003fp} have shown,
one can analyze the form factor dependence on the light quark masses in the
Brodsky-Lepage formula: they concluded that the chiral log does indeed
factorize for the leading term. Nothing is known about subleading terms but
there is no reason to assume that the chiral logs would factorize for them
too. The coefficient of the chiral log in the leading term has been given
explicitly by Chen and Stewart and it is twice as large as what is found at
low energy.  At intermediate energies some sort of transition between the
two regimes must therefore happen, and we see no reason why factorization
should be valid in this region.

In the future, we plan to extend this analysis to heavy-light form factors
which represent one of the most interesting areas of application of \hpc.

\begin{acknowledgments} 
  We thank Heiri Leutwyler for discussions and for reading the manuscript. We thank Thomas Becher, Guido Bell and Simon Aufdereggen for helpful discussions. The Albert Einstein Center for Fundamental Physics is supported by the ``Innovations- und Kooperationsprojekt C-13'' of the ``Schweizerische 
  Universit\"atskonferenz SUK/CRUS''. Partial financial support by the 
  Helmholtz Association through the virtual institute ``Spin and strong
  QCD'' (VH-VI-231) and by the Swiss National Science Foundation is
  gratefully acknowledged. JTC acknowledges a MEC FPU fellowship (Spain).
\end{acknowledgments}
 
 \appendix 

\section{$\pi \pi$ partial waves beyond tree level}
\label{app:partwav}

Consider the dispersive representation of the $\pi \pi$ partial waves
proposed by Roy~\cite{Roy:1971yg}: 
\be
t^I_\ell(s)=k^I_\ell(s)+\sum_{I'=0}^2 \sum_{\ell'=0}^\infty \int_{4
  \mpi^2}^\infty ds' K^{II'}_{\ell \ell'}(s,s') \im t^{I'}_{\ell'}(s') ~,
\label{eq:tIl}
\ee
where $I$ and $\ell$ denote isospin and angular momentum respectively and
$k^{I}_{\ell}(s)$ is the partial wave projection of the subtraction term
\be \label{eq:kIl}
k^I_\ell(s)=a^I_0 \delta_{\ell}^0 + \frac{s-4\mpi^2}{4 \mpi^2} (2 a_0^0-5
a_0^2) \left( \frac{1}{3} \delta_{0}^I \delta_{\ell}^0 + \frac{1}{18}
  \delta_{1}^I \delta_{\ell}^1 -\frac{1}{6} \delta_{2}^I \delta_{\ell}^0
\right)~.
\ee
In order to analyze the possible sources for chiral logarithms, we split
eq.~(\ref{eq:tIl}) into the contributions from the $S$- and $P$-waves, the
higher partial waves and the integral from a cut-off $\Lambda^2$ to infinity:  
\bea
\label{eq:roy}
t^I_\ell(s)=k^I_\ell(s) + t^I_{\ell}(s)_{SP} + d_{\ell}^I(s)~,
\eea
where
\be
\label{eq:SP}
t^I_{\ell}(s)_{SP} = \sum_{I'=0}^2 \sum_{\ell'=0}^1 \int_{4
  \mpi^2}^{\Lambda^2} ds' K^{II'}_{\ell \ell'}(s,s') \im t^{I'}_{\ell'}(s')~, 
\ee
and the so called driving terms $d_{\ell}^I(s)$ are given by
\cite{Ananthanarayan:2000ht}
\be
\label{eq:driving}
d_{\ell}^I(s) =\sum_{I'=0}^2 \sum_{\ell'=2}^\infty
\int_{4\mpi^2}^{\Lambda^2} ds' K^{II'}_{\ell \ell'}(s,s')\, \im
t^{I'}_{\ell'}(s') + 
\sum_{I'=0}^2 \sum_{\ell'=0}^\infty \int_{\Lambda^2}^\infty ds'
K^{II'}_{\ell \ell'}(s,s')\, \im t^{I'}_{\ell'}(s')~. 
\ee 
The expressions of the kernels can be found in
ref.~\cite{Ananthanarayan:2000ht}.  The subtraction term could in principle
contain chiral logarithms but the combination of scattering lengths $2
a_0^0-5a_0^2$ does not have any \cite{Gasser:1983yg}.  As we are going to
show, neither the $S$- and $P$-wave contribution nor the driving terms contain
leading chiral logs either.

In the elastic region unitarity relates the imaginary part of the $\pi \pi$
partial wave to its modulus squared:
\be
\label{eq:optical}
\mathrm{Im}\,t_{l}^I(s)=\sigma (s)\,|t_{l}^I(s)|^2~.
\ee
The integral over the $S$- and $P$-waves may in principle give rise to chiral
logarithms:
\bea
\label{eq:SP_int}
t^I_{\ell}(s)_{SP} &=& \int_{4\mpi^2}^{\Lambda^2} ds' 
K^{I0}_{\ell
  0}(s,s')\left[\frac{1}{2}\,a_0^0b_0^0\,s'
+\frac{1}{16}(b_0^2)^2 (s'^2-8 s' M^2)+\cO(M^4)+\cO(p^6)\right] \\   \nonumber
&+& \int_{4\mpi^2}^{\Lambda^2} ds' K^{I1}_{\ell
  1}(s,s')\left[\frac{1}{16}\left(a_1^1\right)^2(s'^2-8 s'
  M^2)+\cO(M^4)+\cO(p^6)\right] \\ \nonumber  
&+& \int_{4\mpi^2}^{\Lambda^2} ds'
K^{I2}_{\ell
  0}(s,s')\left[\frac{1}{2}a_0^2b_0^2s'
+\frac{1}{16}(b_0^2)^2(s'^2-8 s' M^2)+\cO(M^4)+\cO(p^6)\right]~.
\eea 
The partial wave amplitudes needed for the form factors are $t_{0}^{0}(s)$
for $F_{S}(s)$ and $t_{1}^{1}(s)$ for $F_{V}(s)$. 
Eq.~(\ref{eq:SP_int}) involves integrals of the same type of those
discussed in sec.~\ref{sec:integration}. 
We find that for both partial waves the coefficient of $\ln M^2/s$ is
proportional to 
\be
\left[2 a_0^0 b_0^0 - 5 a_0^2 b_0^2 \right]-
\frac{3 M^2}{4} \left[ 2 (b_0^0)^2+27 (a_1^1)^2-5(b_0^2)^2 \right].
\ee
Although the individual terms in this combination are of order $M^2$, they
cancel and leave as leading contribution something of $\cO(M^4)$ and
therefore beyond the accuracy of this calculation.

According to the expressions of the kernels, terms proportional to $L$ can
be generated only from $\im t(s)$ up to $\cO(p^4)$. Therefore
$d_{\ell}^I(s)$ cannot produce leading chiral logarithms because there $\im
t(s)$ starts at $\cO(p^8)$.  

Since the $\pi\pi$ scattering amplitude at tree level and zero
momentum vanishes linearly in the chiral limit, it does not contain terms 
proportional to $L$. Therefore  
\be
t^{ (2)}(s) = \cO(M^2) + \cO(s)~.
\ee
From unitarity, by applying the chiral power counting to
eq.~(\ref{eq:optical}), it follows that  
\be
\mathrm{Im}\,t^{(4)}= \sigma(s)\,\lvert t^{(2)}\rvert^2 \;\;.
\ee
Hence also the imaginary part of the partial wave to order $p^4$ contains
no term proportional to $L$. Using Roy equations (\ref{eq:roy}), we can
get the corresponding partial wave $t^{(4)}$ from the imaginary part. Since
our previous explicit calculation shows that neither the integration nor the
subtraction term  produce any unwanted chiral logarithms, the partial wave to
next-to-leading order has no terms proportional to $sL$, hence $\beta=0$ in
eq.~(\ref{eq:beta}). By induction one may reach the same conclusion to all
chiral orders provided one does not consider inelastic contributions. This
shows that for the partial waves expanded in $M^2 \ll 
s$: 
\be
t^{(2n)}(s) \!= \bar t^{(2n)}\!(s) + L \sum_{k=0}^{n-1} \beta_k s^k
M^{2(n-1-k)} + \cO(M^{2})  
\ee
the coefficient $\beta_{n-1}$ vanishes to all orders.

\section{Inelastic contributions: explicit expressions of the integrands}
\label{app:inel}
We list here the single contributions to the imaginary part of the scalar form factor from the diagrams $(a)$ to $(g)$ in fig.~(\ref{img:3loop}), namely
\be 
\im F_{\mathrm{inel}}^{(i)}(s)=\frac{1}{2}\int d\Phi_4(s;p_1,p_2,p_3,p_4)\,
Y_{(i)}~.  
\ee
To express the integrands given here below in terms of the variables of our
four-pion phase space parametrization in eq.~(\ref{eq:4pionPS}), we refer
the reader to app.~\ref{app:Lorentz}. The momenta of the final pions are
denoted by $p_e$ and $p_f$, see fig.~\ref{img:4body}, and we define
$s_{ij}:=p_i \cdot p_j$ (notice that $s_{ef}=s/2 - M^2$):
\begin{eqnarray}
{Y_{(a)}}&=&-\frac{5 \left(25 M^2+8 s\right)}{24 F^6} \\
{Y_{(b)}}&=&-\frac{1}{{6 F^6 (M^2-q^2)}}\,\Big\{5 \Big[105 M^4+2 M^2 (55
   (s_{12}+s_{13}+s_{23})+20( s_{14}+ s_{24}+s_{34}) -6 s) \nonumber \\ &&
   +8 \Big(s_{12} (6
   s_{13}+3 s_{14}+6 s_{23}+3
   s_{24} +4 s_{34}-s)
   +s_{23} (4 s_{14}+4 s_{23}+3 s_{24}+3 s_{34}-s)
    \nonumber \\ &&
   +s_{13} (3 s_{14}+6 s_{23}+4 s_{24}+3 s_{34}-s)+4
   s_{12}^2 +4 s_{13}^2 \Big)\Big] \Big\}
    \\
{Y_{(c)}}&=&-\frac{1}{6 F^6
   \,(M^2-q^2)}\,\Big\{5 \Big(3 M^2+2
   (s_{12}+s_{13}+s_{23})\Big) 
   \nonumber \\ &&
   \times \Big[3
   \left(\frac{s}{2}-M^2\right)+10\, M^2+5(s_{12}+s_{13}+s_{23}) +
   3(s_{14}+ s_{24}+s_{34})-s \Big]\Big\} \\ \nonumber
{Y_{(d)}}&=&-\frac{1}{12 F^6
   (M^2-q^2)^2}\Big\{\Big[15 M^4+20 M^2 
   (s_{12}+s_{13}+s_{23})
   \\ \nonumber &&
   + 4 \,\Big( 3 (s_{12}^2 +s_{13}^2+ s_{23}^2)+2
   (s_{13}s_{12}+s_{23}s_{12} + s_{13}s_{23})\Big)\Big] 
    \\ &&
    \times
   (14 M^2+10 (s_{12}+ s_{13}+ s_{23})+ 6 (s_{14}+ s_{24}+ s_{34})+s)\Big\} \\
{Y_{(e)}}&=&\frac{5}{4\, F^6  \nonumber
   [M^2-(p_1+p_2-p_e)^2]} \Big\{2 M^2 (5\, s_{12}+3 (s_{ef}-s_{1e}-s_{2e}-s_{1f}-s_{2f}
   +s_{34})   \\ \nonumber &&
   +4(s_{13}+s_{14}+s_{23}+s_{24}-s_{3e}-s_{4e}-s_{3f}-s_{4f}))
   +4 \Big[s_{12}\, (s_{34}+s_{ef}-s_{1f}-s_{2f}
   \\ \nonumber &&
   +2(s_{13}+s_{14}+s_{23}+s_{24}-s_{3e}-s_{4e}-s_{3f}-s_{4f}))
   -(s_{1e}+ s_{2e}) (s_{13}+ s_{14}+s_{23}
   \\ &&
   +s_{24}+s_{34}+s_{ef}-s_{3e}-s_{4e}-s_{1f} -s_{2f}
   -s_{3f}-s_{4f})\Big]+11\, M^4 \Big\} \\ 
Y_{(f)}&=&
\frac{1}{2 F^6
   [M^2-(p_1+p_3+p_4)^2] (M^2-q^2)}\Big\{\Big[15\, M^4 +10\, M^2 (s_{12}+2
 s_{13} +s_{14}+s_{23}+s_{34} ) \nonumber 
    \\ \nonumber &&
    +4 \Big(s_{13} (s_{12}+3 s_{13}+s_{23})
   +\,s_{14} \,(s_{12}+s_{13}+3
   s_{23})+ s_{34} (3 s_{12}+s_{13}+s_{23}) \,
   \Big)\Big] 
    \\ &&
   \times (3 M^2+2 s_{12}+2 s_{23}+2 s_{24}-3 s)\Big\} \\
{Y_{(g)}}&=&-\frac{1}{2 F^6
   (M^2-q^2)
  [M^2-(p_1+p_2-p_e)^2]}\Big\{\!-55\, M^6 -10 M^4 (10 s_{12}+7 s_{13}+4
s_{14}-3 s_{1e} \nonumber
  \\ \nonumber &&
 -3 s_{1f}+7 s_{23}+4 s_{24}-3 s_{2e}
   -3 s_{2f}+3 s_{34}-4 s_{3e}-4 s_{3f}-4 s_{4e}-4
   s_{4f}+3 s_{ef}) 
   \\ \nonumber &&
   -20\, M^2
   \Big[3 s_{12}^2+ s_{12} (5
   s_{13}+4 s_{14}-s_{1e}-2 s_{1f}+5
   s_{23}+4 s_{24}-s_{2e}-2 s_{2f}+2
   s_{34}-4 s_{3e}
   \\ \nonumber &&
   -4 s_{3f}-4
   s_{4e}-4 s_{4f}+2 s_{ef})
   +s_{13}^2+s_{23}^2-s_{14}
   \,s_{1e}+s_{1e}\, s_{1f}+s_{14}
   \,s_{23}-2 s_{1e}\, s_{23}
   \\ \nonumber &&
   -s_{1f}\,
   s_{23}-s_{1e}\, s_{24}+s_{23}\,
   s_{24}-s_{14} \, s_{2e}+s_{1f}\, 
   s_{2e}
   -2 s_{23}\, s_{2e}
   -s_{24}\,
   s_{2e}+s_{1e}\, s_{2f}-s_{23}\,
   s_{2f}
   \\ \nonumber &&
   +s_{2e} \, s_{2f}-s_{1e} \, s_{34}+s_{23} \, s_{34}-s_{2e} \,
   s_{34}+s_{1e} \, s_{3e}-s_{23}\, s_{3e} +s_{2e} \,s_{3e}+s_{1e} \,
   s_{3f}-s_{23} \, s_{3f} 
   \\ \nonumber &&
   +s_{2e} \,
   s_{3f}+s_{1e} \, s_{4e}
   -s_{23} \,
   s_{4e}+s_{2e}  \, s_{4e}+s_{1e} \,
   s_{4f}-s_{23}  \, s_{4f}+s_{2e} \, 
   s_{4f}-s_{1e}  \, s_{ef} +s_{23} \,  s_{ef}
   \\ \nonumber &&
   -s_{2e}  \, s_{ef}+s_{13} \, 
   (s_{14}-2 s_{1e}-s_{1f}
   +2 s_{23}+s_{24}-2 s_{2e}-s_{2f}+s_{34}-s_{3e}-s_{3f}-s_{4e}-s_{4f}
   \\ \nonumber &&
   +s_{ef})\Big]-8 \Big[3 s_{12}^2 (2 s_{13}+2 s_{14}-s_{1f}+2 s_{23}+2
   s_{24}-s_{2f}+s_{34}-2 s_{3e}-2 s_{3f} -2 s_{4e}
   \\ \nonumber &&
   -2 s_{4f}+s_{ef})
    + s_{12} (s_{13}-s_{1e}+s_{23}-s_{2e})
   (2 s_{13}+2 s_{14}-s_{1f}+2
   s_{23}+2 s_{24}-s_{2f}+s_{34}
   \\ \nonumber &&
   -2 s_{3e}-2 s_{3f}-2 s_{4e}-2 s_{4f}+s_{ef})
   -s_{13}^2\, (2 s_{1e}+s_{2e})
   +s_{13}\, \Big(\!-s_{14} (s_{1e}+2
   s_{2e})
   \\ \nonumber &&
    +s_{2e}\, (2
   s_{1f}-3 s_{23}-2 s_{24}+2
   s_{2f}-2 s_{34}+s_{3e}+2 s_{3f} +2
   s_{4e}+s_{4f}-2 s_{ef})
   \\ \nonumber &&
   +s_{1e}\, (2
   s_{1f}-3 s_{23}-s_{24}+2 s_{2f}-2
   s_{34}+2 s_{3e}+s_{3f} +s_{4e}+2
   s_{4f}-2 s_{ef})\Big)
   \\ \nonumber &&
   +s_{23}\, \Big(\!s_{1e}
   (2 s_{1f}-s_{23}-2 s_{24}+2
   s_{2f} -2 s_{34}+s_{3e}+2 s_{3f}+2
   s_{4e}+s_{4f}-2 s_{ef})
   \\ \nonumber &&
   +s_{2e} (2
   s_{1f}-2 s_{23}-s_{24}+2 s_{2f}-2 s_{34}+2 s_{3e}
   +s_{3f}+s_{4e}+2
   s_{4f}-2 
   s_{ef})\\ &&
   -s_{14} (2  s_{1e}+s_{2e})\Big)\Big]\Big\} 
\end{eqnarray}

\section{Four-pion phase space and related Lorentz transformations}
\label{app:Lorentz}
\begin{figure}[t]
 \centering
 \includegraphics[width=0.45\textwidth]{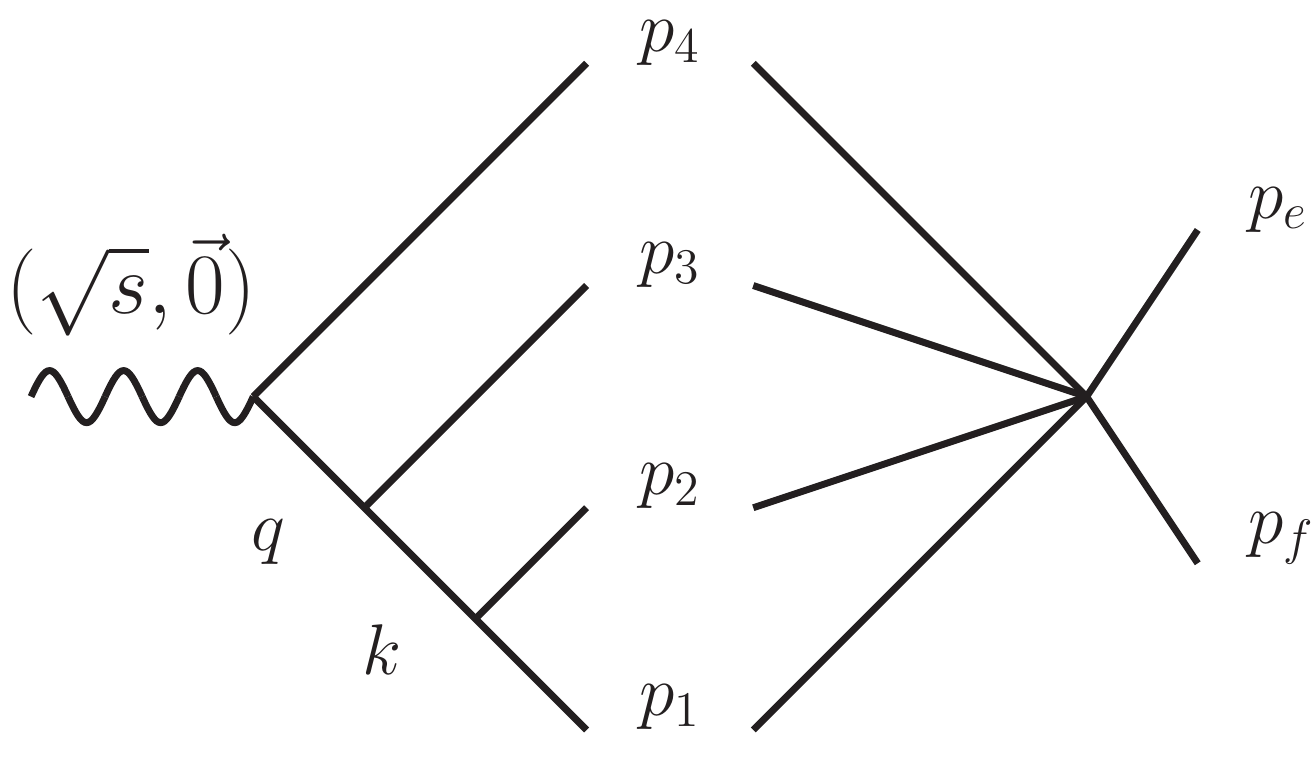}  
 \caption{The kinematic invariants corresponding to our choice of
   parametrization: $s=(p_1+p_2+p_3+p_4)^2$, $q^2=(p_1+p_2+p_3)^2$ and
   $k^2=(p_1+p_2)^2$.} 
 \label{img:4body}
\end{figure}

In order to perform the integral over the four-pion phase space in
eq.~(\ref{eq:4pionPS}) we need to express all the dot products $p_i\cdot
p_j$ entering the Feynman rules for the cut diagrams in terms of our
integration variables. We define angles in the following frames:
$\Sigma_s$, which is the center of mass frame (CMF) of the two final pions,
$\Sigma_q$, which is the CMF of the particles $p_1$, $p_2$ and 
$p_3$, and $\Sigma_k$ which is the CMF of the particles $p_1$and $p_2$. The
labels of the particle momenta are explained in fig.~\ref{img:4body}. 
\begin{figure}[t]
 \centering
 \includegraphics[width=0.8\textwidth]{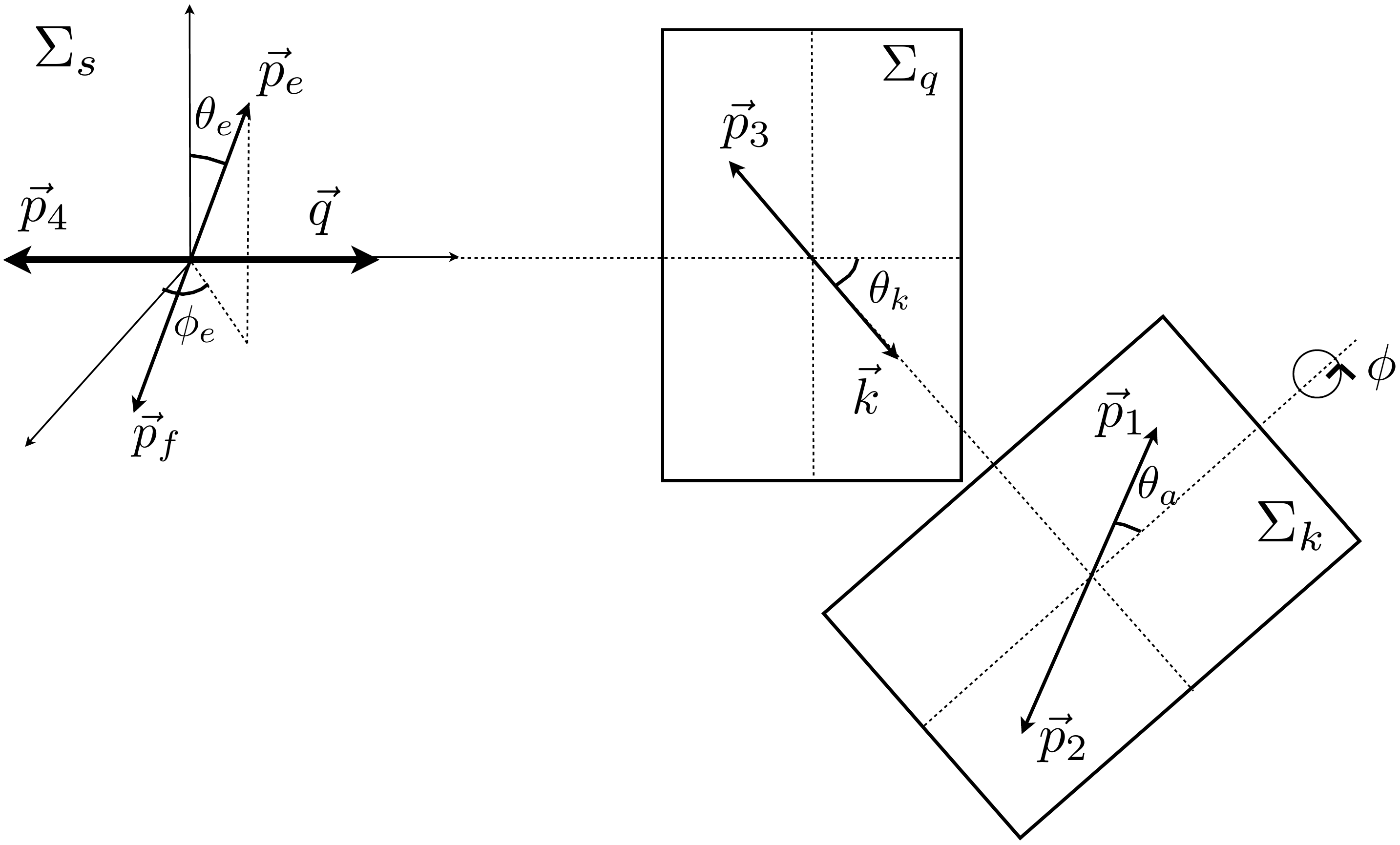}  
 \caption{Our choice of reference frames and angles.}
 \label{img:massi}
\end{figure}
The angles and the corresponding reference frames are shown in
fig.~\ref{img:massi}. The latter ones are connected by the following
Lorentz transformations  
\be
\Sigma_s  \quad \mathop{\leftarrow}\limits^{\Lambda_1}\quad \Sigma_q\quad
\mathop{\leftarrow}\limits^{\Lambda_2} \quad \Sigma_k\,\,. 
\ee
These are given by
\be
\Lambda_1 = \left(
\begin{array}{cccc}
 \frac{s+q^2-M^2}{2 \sqrt{q^2} \sqrt{s}} & 0 & \frac{\lambda^{1/2}
   \left(s,q^2,M^2\right)}{2 \sqrt{q^2} 
   \sqrt{s}} & 0 \\
 0 & 1 & 0 & 0 \\
 \frac{\lambda^{1/2} \left(s,q^2,M^2\right)}{2 \sqrt{q^2} \sqrt{s}} & 0 &
 \frac{s+q^2-M^2}{2 \sqrt{q^2} 
   \sqrt{s}} & 0 \\
 0 & 0 & 0 & 1
\end{array}
\right)
\ee
and by 
\be 
\Lambda_2 = \left(
\begin{array}{cccc}
 \frac{k^2+q^2-M^2}{2 \sqrt{k^2} \sqrt{q^2}} & \frac{\lambda^{1/2}
   \left(k^2,q^2,M^2\right)}{2 
   \sqrt{k^2} \sqrt{q^2}}\sin \phi  & \frac{ \lambda^{1/2}
   \left(k^2,q^2,M^2\right)}{2 \sqrt{k^2} 
   \sqrt{q^2}}\cos \phi  & 0 \\
 0 & \cos \phi  & -\sin \phi  & 0 \\
 \frac{ \lambda^{1/2} \left(k^2,q^2,M^2\right)}{2 \sqrt{k^2} \sqrt{q^2}} \cos \theta _k&
   \frac{k^2+q^2-M^2}{2 \sqrt{k^2} \sqrt{q^2}}\sin \phi  \cos \theta _k &
   \frac{ k^2+q^2-M^2}{2 \sqrt{k^2} \sqrt{q^2}}\cos \phi \cos \theta _k &
   -\sin \theta _k \\
 \frac{\lambda^{1/2} \left(k^2,q^2,M^2\right)}{2 \sqrt{k^2} \sqrt{q^2}} \sin\theta _k &
   \frac{ k^2+q^2-M^2}{2 \sqrt{k^2} \sqrt{q^2}} \sin \phi \sin \theta _k&
   \frac{ k^2+q^2-M^2}{2 \sqrt{k^2} \sqrt{q^2}}\cos \phi  \sin \theta _k&
   \cos \theta _k
\end{array}
\right)\,,
\ee
with the usual K\"all\'en function
$\lambda(x,y,z)=x^2+y^2+z^2-2xy-2xz-2yz$. Now we give the momentum vectors
in the frames where they have the simplest form.  
We define the four-vectors $p_1$ and $p_2$ in the frame $\Sigma_k$ as follows
\be 
p_1''=\frac{\sqrt{k^2}}{2}\left(
\begin{array}{c}
 1 \\
 0 \\
\sigma (k^2) \sin \theta _a \\
 \sigma (k^2) \cos \theta _a
\end{array}
\right)\,\,,\quad
p_2''=\frac{\sqrt{k^2}}{2}\left(
\begin{array}{c}
 1 \\
 0 \\
-\sigma (k^2) \sin \theta _a \\
- \sigma (k^2) \cos \theta _a
\end{array}
\right)\,\,,
\ee
with $\sigma(x)=\sqrt{1-4\mpi^2 /x}$. We give the four-vector $p_3$ in the
frame $\Sigma_q$ 
\be 
p_3'=\left(
\begin{array}{c}
 \frac{q^2-k^2+M^2}{2 \sqrt{q^2}} \\
 0 \\
 -\frac{\lambda^{1/2} \left(q^2,k^2,M^2\right)}{2 \sqrt{q^2}} \cos \theta _k\\
 -\frac{ \lambda^{1/2} \left(q^2,k^2,M^2\right)}{2 \sqrt{q^2}}\sin \theta _k
\end{array}
\right)
\ee
and the four-vectors $p_4$, $p_e$ and $p_f$ in $\Sigma_s$
\bea 
p_4=\left(
\begin{array}{c}
 \frac{s-q^2+M^2}{2 \sqrt{s}} \\
 0 \\
 -\frac{\lambda^{1/2} \left(s,q^2,M^2\right)}{2 \sqrt{s}} \\
 0
\end{array}
\right)\,,\quad &&\\ \nonumber
p_e= \frac{\sqrt{s}}{2}\left(
\begin{array}{c}
1 \\
 \sigma (s) \sin \theta _e \cos \phi _e \\
  \sigma (s) \sin \theta _e \sin\phi _e \\
 \sigma (s) \cos \theta _e
\end{array}
\right)\,,\quad
&&p_f=\frac{\sqrt{s}}{2}\left(
\begin{array}{c}
 1 \\
 - \sigma (s) \cos \phi_e \sin \theta _e \\
 - \sigma (s) \sin \phi_e \sin \theta _e \\
 - \sigma (s) \cos \theta _e
\end{array}
\right)\,.
\eea

After transforming all these momentum vectors onto the same frame, all
Lorentz invariants are written as functions of the seven phase space
variables $k^2$, $q^2$, $\theta_k$, $\theta_a$, $\theta_e$, $\phi$ and
$\phi_e$.

\bibliographystyle{jhep}
\bibliography{factors}

\providecommand{\href}[2]{#2}\begingroup\raggedright\begin{thebibliography}{10}

\bibitem{Weinberg:1978kz}
S.~Weinberg, {\it {Phenomenological Lagrangians}},  {\em Physica} {\bf A96}
  (1979) 327.

\bibitem{Gasser:1983yg}
J.~Gasser and H.~Leutwyler, {\it {Chiral Perturbation Theory to One Loop}},
  {\em Ann. Phys.} {\bf 158} (1984) 142.

\bibitem{Flynn:2008tg}
{\bf RBC Collaboration, UKQCD Collaboration} Collaboration, J.~Flynn and
  C.~Sachrajda, {\it {SU(2) chiral perturbation theory for K(l3) decay
  amplitudes}},  {\em Nucl.Phys.} {\bf B812} (2009) 64--80,
  [\href{http://xxx.lanl.gov/abs/0809.1229}{{\tt arXiv:0809.1229}}].

\bibitem{Bijnens:2009yr}
J.~Bijnens and A.~Celis, {\it {$K \to \pi \pi$ Decays in SU(2) Chiral
  Perturbation Theory}},  {\em Phys.Lett.} {\bf B680} (2009) 466--470,
  [\href{http://xxx.lanl.gov/abs/0906.0302}{{\tt arXiv:0906.0302}}].

\bibitem{Bijnens:2010ws}
J.~Bijnens and I.~Jemos, {\it {Hard Pion Chiral Perturbation Theory for
  $B\to\pi$ and $D\to\pi$ Formfactors}},  {\em Nucl.Phys.} {\bf B840} (2010)
  54--66, [\href{http://xxx.lanl.gov/abs/1006.1197}{{\tt arXiv:1006.1197}}].

\bibitem{Bijnens:2010jg}
J.~Bijnens and I.~Jemos, {\it {Vector Formfactors in Hard Pion Chiral
  Perturbation Theory}},  {\em Nucl. Phys.} {\bf B846} (2011) 145--166,
  [\href{http://xxx.lanl.gov/abs/1011.6531}{{\tt arXiv:1011.6531}}].

\bibitem{Bijnens:2011bp}
J.~Bijnens and I.~Jemos, {\it {Chiral Symmetry and Charmonium Decays to Two
  Pseudoscalars}},  {\em Eur.Phys.J.} {\bf A47} (2011) 137,
  [\href{http://xxx.lanl.gov/abs/1109.5033}{{\tt arXiv:1109.5033}}].

\bibitem{Bijnens:1998fm}
J.~Bijnens, G.~Colangelo, and P.~Talavera, {\it {The vector and scalar form
  factors of the pion to two loops}},  {\em JHEP} {\bf 05} (1998) 014,
  [\href{http://xxx.lanl.gov/abs/hep-ph/9805389}{{\tt hep-ph/9805389}}].

\bibitem{Lepage:1980fj}
G.~P. Lepage and S.~J. Brodsky, {\it {Exclusive Processes in Perturbative
  Quantum Chromodynamics}},  {\em Phys.Rev.} {\bf D22} (1980) 2157.

\bibitem{Chen:2003fp}
J.-W. Chen and I.~W. Stewart, {\it {Model independent results for SU(3)
  violation in light cone distribution functions}},  {\em Phys.Rev.Lett.} {\bf
  92} (2004) 202001, [\href{http://xxx.lanl.gov/abs/hep-ph/0311285}{{\tt
  hep-ph/0311285}}].

\bibitem{Gasser:1990bv}
J.~Gasser and U.~G. Meissner, {\it {Chiral expansion of pion form-factors
  beyond one loop}},  {\em Nucl.Phys.} {\bf B357} (1991) 90--128.

\bibitem{Colangelo:1996hs}
G.~Colangelo, M.~Finkemeier, and R.~Urech, {\it {Tau decays and chiral
  perturbation theory}},  {\em Phys.Rev.} {\bf D54} (1996) 4403--4418,
  [\href{http://xxx.lanl.gov/abs/hep-ph/9604279}{{\tt hep-ph/9604279}}].

\bibitem{Donoghue:1995pd}
J.~F. Donoghue, {\it {On the marriage of chiral perturbation theory and
  dispersion relations}},  \href{http://xxx.lanl.gov/abs/hep-ph/9506205}{{\tt
  hep-ph/9506205}}.

\bibitem{Gasser:1979hf}
J.~Gasser and A.~Zepeda, {\it {APPROACHING THE CHIRAL LIMIT IN QCD}},  {\em
  Nucl.Phys.} {\bf B174} (1980) 445.

\bibitem{Gasser:1980sb}
J.~Gasser, {\it {Hadron Masses and Sigma Commutator in the Light of Chiral
  Perturbation Theory}},  {\em Annals Phys.} {\bf 136} (1981) 62.

\bibitem{Bissegger:2006ix}
M.~Bissegger and A.~Fuhrer, {\it {Chiral logarithms to five loops}},  {\em
  Phys.Lett.} {\bf B646} (2007) 72--79,
  [\href{http://xxx.lanl.gov/abs/hep-ph/0612096}{{\tt hep-ph/0612096}}].

\bibitem{Beringer:1900zz}
{\bf Particle Data Group} Collaboration, J.~Beringer {\em et.~al.}, {\it
  {Review of Particle Physics (RPP)}},  {\em Phys.Rev.} {\bf D86} (2012)
  010001.

\bibitem{Guo:2011ir}
F.-K. Guo, B.~Kubis, and A.~Wirzba, {\it {Anomalous decays of eta' and eta into
  four pions}},  {\em Phys.Rev.} {\bf D85} (2012) 014014,
  [\href{http://xxx.lanl.gov/abs/1111.5949}{{\tt arXiv:1111.5949}}].

\bibitem{Efremov:1979qk}
A.~Efremov and A.~Radyushkin, {\it {Factorization and Asymptotical Behavior of
  Pion Form-Factor in QCD}},  {\em Phys.Lett.} {\bf B94} (1980) 245--250.

\bibitem{Chernyak:1983ej}
V.~Chernyak and A.~Zhitnitsky, {\it {Asymptotic Behavior of Exclusive Processes
  in QCD}},  {\em Phys.Rept.} {\bf 112} (1984) 173.

\bibitem{Bauer:2002nz}
C.~W. Bauer, S.~Fleming, D.~Pirjol, I.~Z. Rothstein, and I.~W. Stewart, {\it
  {Hard scattering factorization from effective field theory}},  {\em
  Phys.Rev.} {\bf D66} (2002) 014017,
  [\href{http://xxx.lanl.gov/abs/hep-ph/0202088}{{\tt hep-ph/0202088}}].

\bibitem{Roy:1971yg}
S.~Roy, {\it {Exact integral equation for pion-pion scattering involving only
  physical region partial waves}},  {\em Phys. Lett. B} {\bf 36} (1971) 353.

\bibitem{Ananthanarayan:2000ht}
B.~Ananthanarayan, G.~Colangelo, J.~Gasser, and H.~Leutwyler, {\it {Roy
  equation analysis of $\pi \pi$ scattering}},  {\em Phys.Rept.} {\bf 353}
  (2001) 207--279, [\href{http://xxx.lanl.gov/abs/hep-ph/0005297}{{\tt
  hep-ph/0005297}}].

\end{thebibliography}\endgroup


\end{document}